# The development of a testbed for the X-ray Interferometer mission

R. den Hartog[1], P. Uttley[2], R. Willingale[3], H. Hoevers[1], J.-W. den Herder[1], M. Wise[1]

[1] SRON Netherlands Institute for Space Research, Utrecht, The Netherlands
[2] University of Amsterdam, The Netherlands
[3] University of Leicester, United Kingdom

## ABSTRACT

An X-ray Interferometer (XRI) has recently been proposed as a theme for ESA's Voyage 2050 planning cycle, with the eventual goal to observe the X-ray sky with an unprecedented angular resolution better than 1 micro arcsec (5 prad) [1]. A scientifically very interesting mission is possible on the basis of a single spacecraft [2], owing to the compact 'telephoto' design proposed earlier by Willingale [3]. Between the practical demonstration of X-ray interferometry at 1 keV by Cash et al. [4] with a 1 mm baseline and 0.1 arcsec effective resolution to a mission flying an interferometer with a baseline of one or more meters, an effective collecting area of square meters and micro arcsec resolution lie many milestones. The first important steps to scale up from a laboratory experiment to a viable mission concept will have to be taken on a scalable and flexible testbed set-up. Such a testbed cannot singularly focus on the optical aspects, but should simultaneously address the thermal and mechanical stability of the interferometer. A particular challenge is the coherent X-ray source, which should provide a wavefront at the entrance of the interferometer that is transversely coherent over a distance at least equal to the baseline, and bright enough.

In this paper, we will explore the build-up of a testbed in several stages, with increasing requirements on optical quality and associated thermo-mechanical control and source sophistication, with the intent to guide the technological development of X-ray interferometry from the lab to space in a sequence of achievable milestones.

## 1. INTRODUCTION

For a potential future ESA mission based on an XRI (X-ray interferometer) [1] and [2], we are developing a series of testbeds to experiment with, develop the enabling technology on, and eventually demonstrate the capabilities of an X-ray interferometry platform. The concept for this grazing-incidence interferometer is based on a design by Willingale [3] and is shown in Figure 1 below. It consists of a configuration of 4 flat mirrors, in which the M2 is slatted (like half-opened Venetian blinds). It is extremely compact by making optimal use of the ratio of the Bragg reflection angle on M2 and M4 and the beam angle required for interferometry after M2 (exemplified in Figure 1c). At the expense of an extra reflection, the size of the optical constellation can be brought down from the focal length (depending on desired angular resolution several 10's to 10's of thousands km, see e.g. [5] and [6]). An optical test of the concept was already performed in 2005 [7], but a proper X-ray demonstration has never taken place. The series of tests we propose would be an extension of the experiments at Bessy by Leitenberger et al. [8] (40 μm baseline at 7 keV) and by Cash et al. [4] in the US (1 mm baseline at 1.25 keV).

A key challenge in testing this interferometer is to construct an X-ray source that mimics an astrophysical source as closely as possible. In contrast with a monochromatized beam (as usually provided by a synchrotron or electroluminescence), an astrophysical source has a low longitudinal coherence length, $l_c^{(l)} = \lambda R$, with $R = E/\Delta E \cong 100$ given by the X-ray energy resolving power of the (semi-conductor) detector, and a large transverse coherence length, due to its vast distance. Instant interferometry for photon energies in the range 1

– 7 keV requires sufficient conservation of longitudinal (temporal) coherence, i.e. the path length along the alternative routes a photon can take through the interferometer needs to remain stable during an integration time sample to within a coherence length in order for the fringe pattern that constitutes the astronomical record not to get smeared out. Real unresolved astronomical sources produce X-rays that are plane-parallel and transversely coherent, but not longitudinally coherent, due to their broadband X-ray emission. However, the use of energy resolving detectors restores the longitudinal coherence length to useful dimensions.

The XRI will detect energies in the range 1 – 7 keV, which correspond to wavelengths in the range 0.17 – 1.2 nm. Modern semi-conductor based detectors, e.g. the pnCCD X-ray Color Camera (XCC) [9] or the DEPFET-based APS camera on the Athena WFI instrument [10], offer energy resolutions $\Delta E$ below 100 eV at 1 keV and 145 -170 eV at 7 keV, bringing the coherence length $l_c$ to 7.5 nm at 1 keV and 3.5 nm at 7 keV. Cryogenic detectors, such as Transition Edge Sensors [11], could improve $\Delta E$, and thus $l_c$ by a factor 50 – 100, but pose such a greater challenge in collecting area and complexity of the cryostat technology that we refrain from considering them in this initial study.

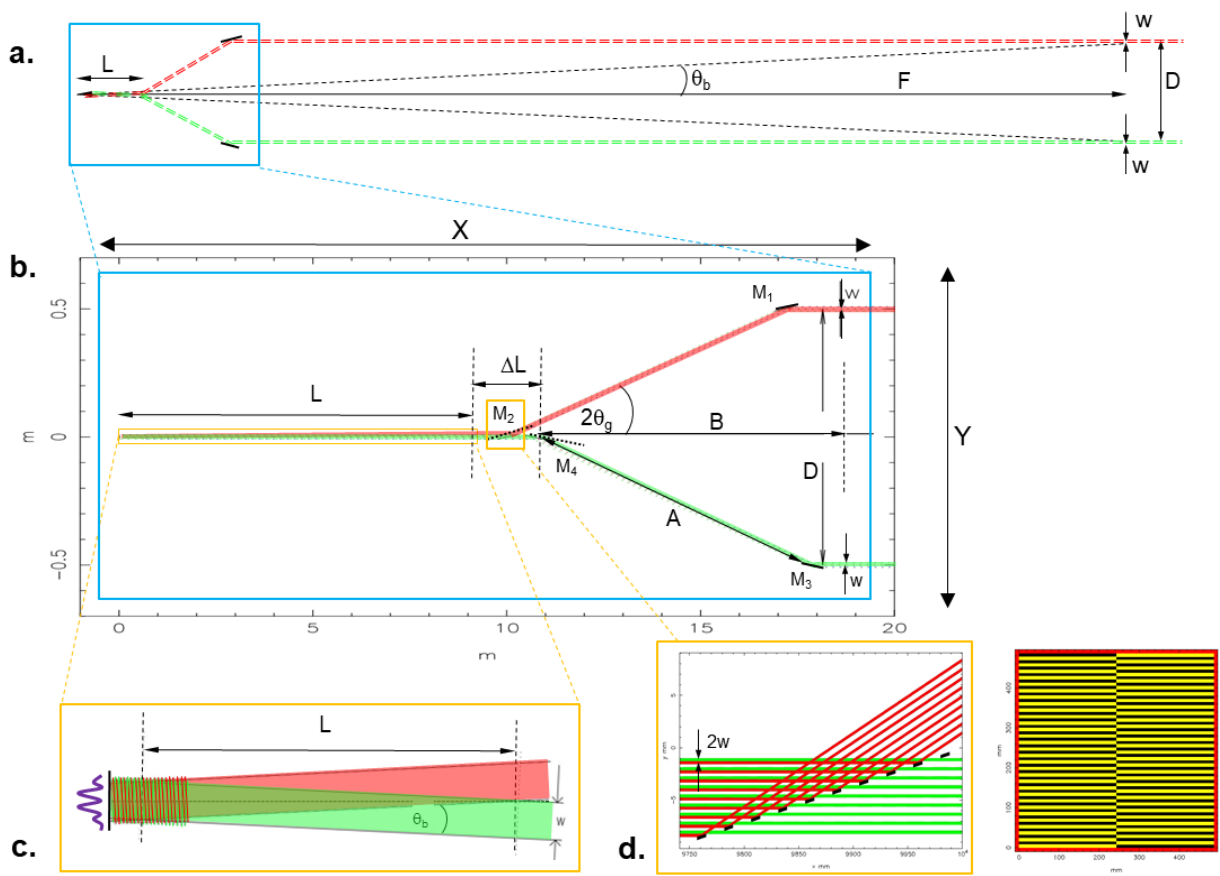

Figure 1. Schematic representation of the optics of the X-ray interferometer, at various levels of detail. a. Overview, comparing the classical and the telephoto geometry. Here D is the effective baseline, F the focal length, w the width of a single beam and L the length over which the interference of X-ray beams takes place. b. Zoom in on the telephoto geometry, showing how flat mirrors M1, M3 and M4 and slatted mirror/beam-combiner M2, plus the detector plane fit inside a compact volume, defined by dimensions X and Y, which is roughly compatible with the dimensions of the largest Ariane 6 launcher fairing. c. Zoom in on the interference of one pair of beams, explaining the relation for the beam angle $\theta_b$ = w/L. d. Side view and front view on the slatted mirror/beam-combiner M2.

Thermo-mechanical disturbances of the system during observations will introduce pathlength variations δx along the alternative photon routes to the detector, and will thus reduce the quality of the recorded interferogram. This quality is usually expressed as fringe visibility $V = (I_{max} - I_{min})/(I_{max} + I_{min})$, with $I_{max}$ and $I_{min}$ the intensities at the maximum and minimum of the fringe.

This paper is organized as follows: We first recapitulate the essential details of the telephoto geometry, and the way in which we optimize it to the boundary conditions, then we analyze the sensitivity of the fringe visibility to the most common geometrical disturbances, and the prospects of realizing an structure that offer sufficient thermo-mechanical stability. Finally we discuss the experimental set-ups needed for testing the realization of these requirements as a combination of coherent X-ray source and testbed.

## 2. THE TELEPHOTO GEOMETRY

Figure 1 shows the Willingale telephoto design [3], consisting of two flat mirror pairs M1 - M2 and M3 - M4. By making optimal use of the ratio between the grazing angle and beam angle the interferometer can be compactified by factors of several $10^3$.

The following points are driving the design and capabilities of the X-ray interferometer:
- The interferometer has to operate in the energy range of 1 to 7 keV for astrophysical reasons (see [1] and [2]). Together with the desired baseline, $D$, this determines the angular resolution $\Delta\theta = \frac{1}{2}\lambda/D$, which is the main performance parameter.
- The length of the interferometer (parameter $X$ in Figure 1b) must be compatible with the host structure. For instance, for a single spacecraft mission, $X$ must be compatible with a launch on an Ariane A64 / A62, which has a 20 m long fairing, with a diameter of 5.4 m [12], and for a testbed, it needs to fit e.g. inside the Panter vacuum tank [13] or the available space in a synchrotron facility (TBD).
- There is a physical limit on the size of the effective pixels on a detector, $\Delta p$. The smaller the pixel, the compacter the design can be. The effective pixel size is defined here as the spatial resolution with which the location of an incoming photon in the direction across the fringes can be determined. At maximum, $\Delta p$ is half the fringe separation $\Delta f$, but ideally more pixels per fringe are available.
- Since only flat mirrors are used, there is no optical magnification. This implies that the collecting area of the telescope is determined by the number of pixels in the image plane.
- There is a physical limit on the grazing incidence angle, $\theta_g$, as shown in Table 1 for three commonly used materials, which is strongly dependent on photon energy.

Table 1. Theoretical reflectivity R of mirrors at different X-ray energy and incidence angles, 0.1 nm $\cong (\lambda_{1keV}/14, \lambda_{7keV}/2)$ rms roughness, obtained from [14].

| Grazing angle $\theta_g$ | SiC | | Pt | | Ir | |
|---|---|---|---|---|---|---|
| | 1 keV | 7 keV | 1 keV | 7 keV | 1 keV | 7 keV |
| 2.0° | 0.54 | 3.1 $10^{-5}$ | 0.62 | 8.7 $10^{-4}$ | 0.65 | 9.6 $10^{-4}$ |
| 1.5° | 0.83 | 1.0 $10^{-4}$ | 0.72 | 3.1 $10^{-3}$ | 0.73 | 3.4 $10^{-3}$ |
| 1.0° | 0.91 | 5.6 $10^{-4}$ | 0.81 | 2.3 $10^{-2}$ | 0.82 | 2.5 $10^{-2}$ |
| 0.75° | 0.93 | 1.9 $10^{-3}$ | 0.85 | 0.15 | 0.86 | 0.17 |
| 0.5° | 0.96 | 1.3 $10^{-2}$ | 0.90 | 0.77 | 0.91 | 0.79 |
| 0.25° | 0.98 | 0.94 | 0.95 | 0.91 | 0.95 | 0.91 |

The top-level design flow is thus as follows:
- The minimum allowed fringe separation is given by the pixel size and the number of pixels per fringe (ppf): $\Delta f = \Delta p \cdot \text{ppf}$

- Together with the wavelength $\lambda$, this gives the beam angle: $\theta_b = \lambda/\Delta f$ (see [3] for a full description of all the geometric relations). Note that without the telephoto design, $X$ would need to be as large as the focal length $F = D / \theta_b$, which is several tens of km.
  - We now try to fit the total interferometer length $L_{tot} = L + B + \Delta L$ (see Figure 1b and c for definitions) inside the maximum space available, $X$.
  - In this relation $B = \frac{1}{2}D/\tan(2\theta_g)$, the length of the interferometer arm A, projected onto the optical axis. The grazing angles $\theta_g$ for $\lambda = 1$ and 7 keV are chosen here as indicated in Table 1 by the shaded fields.
  - $\Delta L$ is the space needed to place mirrors M2 and M4 in line after one another, taken to be one mirror size, 100 mm, in order to fill the aperture optimally with fringes.
  - Now the length of the overlapping beams, $L$, is given by the width of the beam, $w$, and $\theta_b$ as $L = w/\theta_b$ (see Figure 1c). The width of the beam is given by the number of fringes $N_f$ we wish to see and the fringe separation: $w = \Delta f \cdot N_f$. The total number of fringes expected to be visible is $\approx E/\Delta E$. For modern energy resolving X-ray cameras $E/\Delta E > 10$ for energies > 1 keV, hence we start with $N_f = 10$ to find a value for $L$ [9], [10].
  - Note that $w$ defines one slat width on the M2 mirror. The actual mirror sizes, and incoming beams, are going to be defined by multiples of these sub-beams, in order to achieve the required collecting area. At this point in the analysis, however, we consider for simplicity only a single sub-beam.
- Now if $L_{tot} = L + B + \Delta L \leq X$ we have a valid configuration. If $L_{tot} > X$, we work the above analysis back starting with $L = X - (B + \Delta L)$, to find the corresponding $N_f$. If $N_f \geq 2$, we have a valid configuration, otherwise we have to consider changing more parameters.
- With the value for $w$ we also know the minimum size of the mirrors: $s_m = w / \sin(\theta_g)$.

Table 2 lists the outcome of this design exercise for two options of a single spacecraft XRI: one option with a 1 keV baseline of 1 m, another with a baseline of 2 m, offering angular resolutions of resp. 128 and 64 µas.

Table 2. Summary of design parameters for two options of a telephoto optical geometry intended for a single spacecraft design for the XRI.

| parameter | symbol | unit | option 1 | | option 2 | |
|---|---|---|---|---|---|---|
| photon energy | $E$ | | 1 keV | 7 keV | 1 keV | 7 keV |
| total length | $X$ | m | 18.5 | 18.5 | 18.5 | 18.5 |
| baseline | $D$ | m | 1 | 0.25 | 2 | 0.5 |
| equiv. angular res. | $\Delta\theta$ | µas | 128 | 73 | 64 | 36.5 |
| photon spatial resolution | $\Delta p$ | µm | 10 | 10 | 10 | 10 |
| # pixels per fringe | ppf | | 4 | 2 | 4 | 2 |
| grazing angle | $\theta_g$ | mrad | 34.9 | 8.7 | 34.9 | 8.7 |
| min. fringe spacing | $\Delta f$ | µm | 40 | 20 | 40 | 20 |
| beam angle | $\theta_b$ | µrad | 31 | 8.9 | 31 | 8.9 |
| beam width | $w$ | µm | 349 | 99.5 | 127 | 36.1 |
| # visible fringes | $N_f$ | | 8.7 | 5.0 | 3.2 | 1.8 |
| beam overlap length | $L$ | m | 11.25 | 11.24 | 4.10 | 4.08 |
| projected arm length | $B$ | m | 7.15 | 7.16 | 14.30 | 14.32 |
| focal length | $F$ | km | 32.3 | 28.2 | 64.5 | 56.5 |
| min. mirror size | $s_m$ | mm | 10.0 | 11.4 | 3.6 | 4.1 |

The following design parameters were used:
- The total length $X$ was chosen to be 18.5 m, which is well shorter than the length inside an A64 fairing, but might still turn out to be a too optimistic value when eventually all system aspects are taken into consideration. $X$ is not a hard limit: it could easily be reduced by a couple of meters with only a modest penalty on required $\Delta p$ or $N_f$.

- The effective photon spatial resolution Δ*p* was taken equal to 10 μm, which is slightly more optimistic than the 12 μm sub-pixel resolution achieved with the XCC.
- The baselines for the 7 keV photons are necessarily a factor 4 smaller, because of the smaller grazing angle at which still an acceptable reflectivity is achieved for 7 keV photons (see Table 1).

This very basic design exercise high-lights already a number of trades that will need to be made towards a full mission design:
- There is freedom to choose larger baselines, and hence better angular resolutions. Even with a 2 m baseline, the number of fringes that can be detected at 1 keV is probably still sufficient: 3 fringes, spread out over 12 pixels. On the other hand, an 0.5 m baseline for 7 keV is already beyond the practical limit: the number of detectable fringes is already below two and those are spread out on 4 pixels. This is hardly a useful interferogram. This situation might be salvaged a bit by tilting the camera, so that the interferogram is sampled by more pixels, but apart from the practical feasibility of such a tilt, it does not increase the marginal number of fringes in the beam.
- Another net effect of a larger baseline is a more than proportionally smaller beam width *w*. Together with the beam height *h*, which is the dimension of the detector array in the direction perpendicular to the fringes, *wh* forms the area of a single slat. There is thus freedom in the dimensioning the slats, and the number of slats in the M2 mirror that can be exploited to further optimize the system design.

## 3. OPTICAL STABILITY REQUIREMENTS

The relation between a pathlength disturbance δx and reduction of fringe visibility V is simulated for a basic double-slit interferometer in Figure 2 below. Compared to the telephoto design in Figure 1, this simulation assumes no M2 and M4 mirrors and an *F* = 30 and 20 km for resp. 1 and 7 keV, rather than a total length X of ~18.5 m. But from the point of view of OPD disturbances due to mirror displacement or rotation, this basic double-slit interferometer is equivalent to the telephoto design, although details of the interference pattern are different: with the slatted M2 mirror the response is always going to be in the near-field regime (see [3]) and the interference fringe profile should be calculated using Fresnel integrals. For the calculations here, the far-field approximation is adequate enough. For both energies considered, the 1 σ OPD variations at the level of δ*x* = λ/14 result in a ~10% reduction of the fringe visibility. This number includes the additional reduction in visibility due to a finite energy resolution of the detector, which is assumed to be 80 eV FWHM at 1 keV and 160 eV at 7 keV (cf. [9]). An identical analysis for 1 σ beam angle variations at the level of δθ equivalent to Δ*f*/14 yields identical results, as would be expected given the geometric relation between δθ and δ*x*.

Figure 2a and b show single pairs of mirrors, where the red mirror is the displaced / rotated version of the black, which is the outer mirror (M1 or M3) of either one of the two pairs in Figure 1. Here we perform a limited sensitivity analysis, equivalent to that in [15], but only for the most critical displacements and rotation.

The displacement of the M1 or M3 mirror in Figure 2a gives the same change in optical path as a similar displacement in opposite direction on M2 or M4. And whereas the effect of a displacement of the whole mirror (a combination of δ*x* or δ*y*) can be seen in time, resulting in a rms fluctuation of the visibility of the fringe created by the four mirrors on the detector plane, it can also be considered as a displacement in space δ*s*, a step height on the mirror, which results in a rms fluctuation of the visibility in time as different incoming photons hit different parts of the mirror. These offsets can be treated on equal footing as the relation between δ*s*, δ*x* and δ*y* is given by: δ*x* = δ*s* / sin $\theta_g$ and δ*y* = δ*s* / cos $\theta_g$. In [3] the mirror quality is expressed in terms of an rms roughness, $\sigma_m$, and treats the problem with first-order perturbation theory in terms of a Total Integrated Scatter, TIS, on two subsequent mirrors, which is to be smaller than 0.1:

$$\sigma_m = \frac{\lambda}{4\pi \sin \theta_g} \sqrt{TIS/2}$$

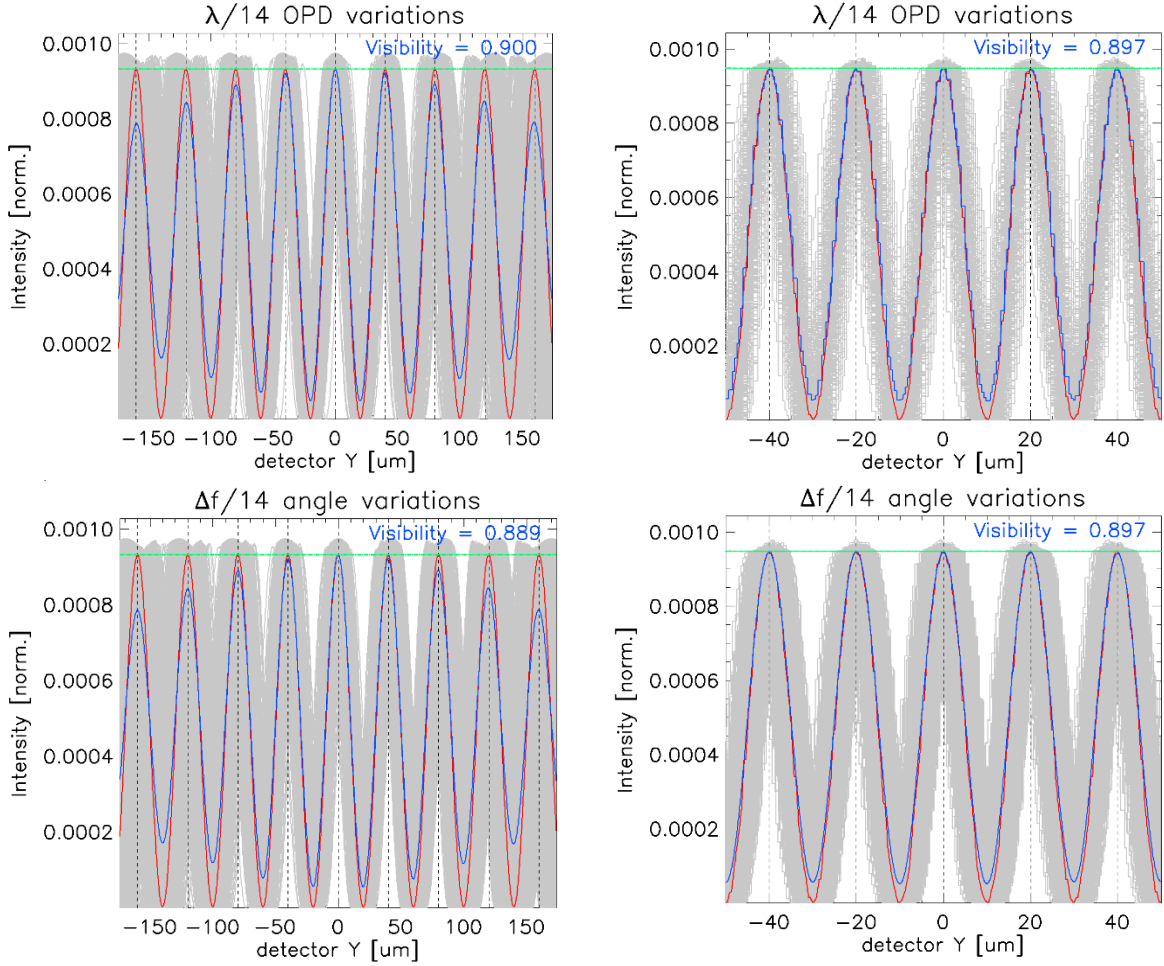

Figure 2. Impact of OPD (top panels) and beam angle variations (bottom panels) on the visibility of fringes at 1 keV (left panels) and 7 keV (right panels), when the pixel energy resolution $\Delta E$ is taken into account as well. The red pattern is the original, mono-chromatic, undisturbed pattern. The grey curves are individual fringes computed for 1000 random realizations in which normal distributed OPD offsets with a 1 $\sigma = \lambda_0/14$ (top) or beam angle offsets equal to $\Delta f/14$ and normal distributed wavelengths $\lambda_i$ with 1 $\sigma = \Delta E_{FWHM,i} / 2.355$. The blue fringes are the result of averaging these 1000 realizations. The simulations are made for a basic double-slit interferometer, equivalent to configuration option 1 described in Table 2. The green line indicates the fringe envelope defined by the finite size $w$ of beam / slit. The vertical lines indicate the location of the fringes, spaced at $\Delta f$, which is twice the maximum width of the detector pixels.

TIS is also a loss of visibility, but one that works in the same way as the finite reflectivity of the mirrors under grazing conditions. In Figure 3a, the difference in OPD due to the displacement of the mirror is equal to $h - \Delta x_1$, given that the alternative pathlengths between lines of equal phase $\phi_0$ and $\phi_1$ are equal. A small contribution $\Delta x_2$ due to the fact that the beam makes an angle $\frac{1}{2}\theta_b$ with the plane of the detector is ignored here. By symmetry, $h = \delta x$. Hence, OPD $= h - \Delta x_1 = \delta x \, (1 - \cos 2\theta_g) = \delta x \, [2 \sin^2 \theta_g]$.

Since we have not performed a system analysis yet, it is too early to define a total visibility requirement and an associated budget for individual contributions. Image reconstruction is also very well possible with a visibility of 50%, as long as it can be properly calibrated; a high visibility is therefore desirable, as it implicates a high efficiency of the optical system, but not a *sine qua non*. In the present analysis we therefore consider the 90% visibility limit per item, and not as a total goal, to obtain an indication for the levels of accuracy and stability that must be met. The maximum allowed mirror offsets $\delta x$ and $\delta y$ are thus given by:

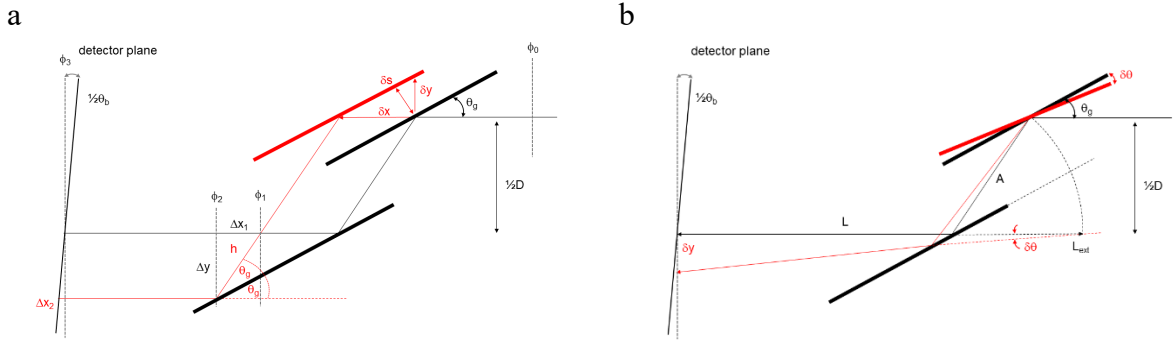

Figure 3. Schematic of ray tracing for sensitivity of the outer mirrors to translation in a. and rotation in b. Note the difference in scale in the X and Y direction, which is even further exaggerated in this Figure. The vertical dashed lines in a. indicate planes of equal phase for photons following the original path or the disturbed path.

$$\delta x < \frac{\lambda_x}{14}\left[2\sin^2\theta_g\right]^{-1}$$

$$\delta y < \frac{\lambda_x}{14}\left[2\sin^2\theta_g\right]^{-1}\tan\theta_g \cong \frac{1}{28}\frac{\lambda_x}{\sin\theta_g}$$

A rotation of one of the mirrors has two effects: it shifts one beam with respect to the other, and thus creates a shift of the fringe pattern on the detector plane (see Figure 3b). Due to that shift the effective arm is also slightly extended by a factor $1/\cos\frac{1}{2}\theta_b$ due to the angle $\frac{1}{2}\theta_b$ between the equal phase planes and the detector plane:

$$\delta\theta < \frac{\Delta f}{14}\left[\frac{\cos\frac{1}{2}\theta_b}{L + \Delta L + A}\right] \cong \frac{1}{14}\frac{\Delta f}{X}$$

The angular stability is associated with a relative stability of the outer ends of the mirror $\delta m = s_m\,\delta\theta$. Table 3 shows the resulting requirements that would have to be met at spacecraft level for both options in Table 2. Increasing (or decreasing) the baseline does not have an impact on the required stability of the mirrors. This has also consequences for the testbed we are planning. In particular, the stability of the mirrors in the direction perpendicular to the beams is required to the nm level. For comparison, the lattice constant of Ir or Pt is 0.38 or 0.39 nm, respectively. Mirrors with this quality are routinely produced by industry, e.g. [16], and a prototype of a slatted mirror has already been produced with nm scale surface quality [17]. A mounting that allows high precision alignment over a larger mirror surface and does not spoil the optical quality may require some additional development activity.

Table 3. Summary of stability requirements for the two options of a telephoto optical geometry intended for a single spacecraft design for the XRI. For each item the indicated rms value corresponds to a 10% loss in visibility.

| parameter | symbol | unit | option 1 | | option 2 | |
|---|---|---|---|---|---|---|
| photon energy | $E$ | | 1 keV | 7 keV | 1 keV | 7 keV |
| min. mirror size | $s_m$ | mm | 10.0 | 11.4 | 3.6 | 4.1 |
| max. mirror offset | $\delta x$ | nm | 36.3 | 83.3 | 36.3 | 83.3 |
| max. mirror offset | $\delta y$ | nm | 1.27 | 0.73 | 1.27 | 0.73 |
| max. rms mirror roughness | $\sigma_m$ | nm | 0.75 | 0.43 | 0.75 | 0.43 |
| max. rotation of mirror | $\delta\theta$ | nrad | 154.3 | 88.1 | 154.1 | 77.2 |
| max. relative offset | $\delta m$ | nm | 1.54 | 0.88 | 0.56 | 0.32 |

# 4. THERMO-MECHANICAL STABILITY REQUIREMENTS

Although operating in a different wavelength range, the GAIA mission offers an interesting heritage in stable opto-thermo-mechanical design [18]. In particular, the requirements of 4 µas uncertainty in star position measurement, and an 0.5 µas (2.5 prad) uncertainty in the measurements of the basic angle variations, demand considerably higher stability during measurements than an XRI: a 2.5 prad described as displacement is a movement of 2.5 pm at a distance of 1 m. The demands of an XRI in the most critical direction (~ 1 nm parallel to the baseline) appear therefore to be two orders of magnitude *less stringent* over similar distances.

The extreme GAIA stability has been achieved in various steps, most of which are also applicable to an XRI:
- Send the spacecraft to a location with extremely stable environmental conditions, i.e. the Earth-Sun Lagrange points L1 or L2. This is the destination of choice of many high-precision missions. The higher cosmic ray background, which affects in particular the X-ray detector, can be tackled by similar techniques as applied in the Athena WFI [10].
- Build the instrument platform of an extremely stable material. In the case of the GAIA basic angle monitor (BAM), Silicon Carbide (SiC) has been chosen as it combines excellent mechanical and thermal properties, in particular a high specific stiffness and a low thermal expansion coefficient.
- Realize a stable thermo-mechanical design of the instrumentation.
- Apply a metrological system in the BAM, stable enough to measure basic angle variations of ±0.5 µas. The BAM is based on the measurement of the relative position of two interferometric patterns, set up via a set of mirrors on the BAM structure, where each pattern is being generated from a common 532 nm laser diode source. GAIA's variations in arm length turned out to be extremely smooth, and there is no reason to expect another behavior for an XRI. Shifts of fringe patterns would then be easy to monitor and correct.
- At room temperature, the CTE of high-quality SiC is $2 \cdot 10^{-6}$ / K (about 6 times lower than stainless steel). In order to keep mirrors in position down to 1.3 nm over a distance of a meter (a requirement on the $\delta y$ of M1 and M3) the temperature throughout the structure needs to remain stable to 0.6 mK. The challenge along the beam direction is more benign: an OPD stability of ~40 nm over ~10 m requires a uniformity of temperature in the structure of ~2 mK (where the beams overlap temperature changes do not result in OPD offsets between the two interfering beams).
- A 1 part in $2 \cdot 10^3$ level of control is challenging, but certainly not impossible. It can even be relaxed by a factor 10 by cooling the SiC to 100 K, an approach taken partially at GAIA (which has an onboard temperature of ~163K). It remains to be seen in a detailed system study whether cooling the platform in addition to thermal control is actually needed for an XRI.
- The final step in reaching the extreme precision of GAIA makes use of the fact that the spacecraft rotates, which causes a periodicity in the output of the BAM signal. By making a fit to this periodic signal it can be removed and the residuals are at the µas level [19]. Rotation is considered an option for the XRI spacecraft to fill the *u,v* plane and create high imaging quality, but needs to be studied on a system level.

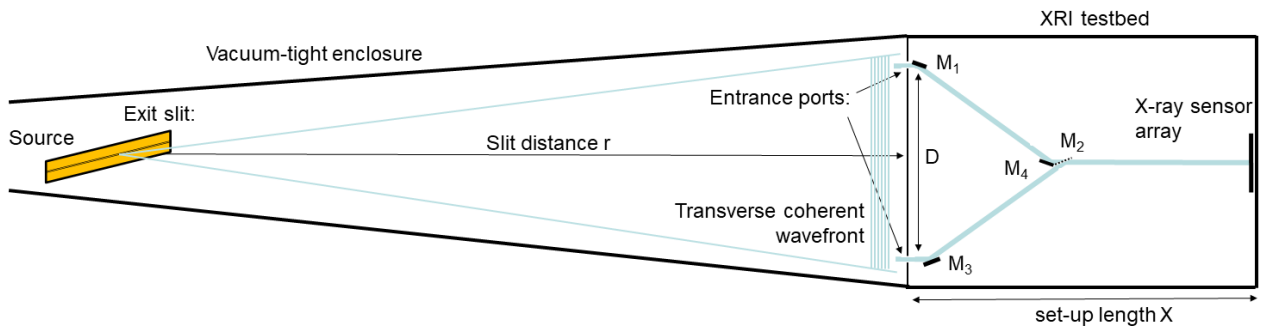

Figure 4. Conceptual sketch of an implementation of the testbed at an X-ray beamline.

It is reassuring to know that a mission with geometrical stability requirements which are two orders of magnitude more stringent than ours, is successfully operating in space today. The current success of the GAIA mission gives credit to the chosen approach. The combination of material choice, thermo-mechanical design and optical metrology allows versatile solutions to reach the required extreme precision and has proven itself in flight. This inspires confidence that the more modest requirements of an XRI will also be met, re-using existing technology.

## 5. DEVELOPMENT OF AN INTERFEROMETRY TESTBED

Between the baseline demonstrated by Cash et al. [4] and that required in the single spacecraft concept discussed above gape three orders of magnitude. We seek to bridge this gap in three subsequent order of magnitude steps with testbeds that demonstrate interferometric fringes for baselines of respectively 10, 100 and 1000 mm. The development of such a series of interferometry testbeds requires a parallel development of X-ray sources, capable of delivering at the entrance of the testbed a wave front that is transverse coherent over at least the distance between the entrance ports. This concept is illustrated in Figure 4.

### 5.1 The source

Our idea to approximate such a source is to make use of increasingly narrow slits in a sufficiently thick (photon stopping, high-Z) material. With sufficient distance $r$ between slit, with diameter $d$, and entrance ports of the interferometer, if $r \gg d^2/\lambda$, the Fraunhofer condition applies. As the transverse coherent part of the (Fraunhofer) diffraction pattern is formed only in the central ~50% of the width of the central peak, the slit width $d$ is related to distance $r$ between the XRI entrance slits and the baseline $D$ (the distance over which the X-ray diffraction pattern needs to be transverse coherent) as $d = \lambda r/2D$. There are two approaches to such a source:

1. A soft X-ray synchrotron beamline. Such beams have a large photon flux, which ensures a fast build-up of the interference pattern, cutting off the frequency bandwidth of instabilities at a higher frequency. However, $r$ larger than a few 10's of m are difficult to realize inside these beamlines, placing heavier requirements on the slit fabrication: e.g. a demonstrations at 7 keV ($\lambda$ = 0.18 nm) for an $r$ of 20 m and a $D$ of 0.25 m would require a slit width of 7 nm in a sufficiently thick substrate (e.g. for a 7 keV transmission below 10%, at least 6 μm of a high-Z material like Ta is needed, implying a 1:1000 aspect ratio).
2. A long X-ray beamline like the MPE Panter, which offers an $r$ of 130 m, places less severe requirements on the slit width. With a pipe diameter of 1 m and a vacuum tank at the end of the pipe of 12 m deep it offers adequate space for the largest version of the testbed. The challenge here is in the photon flux. Although the electro-luminescent sources are capable of delivering a uniform flux of $10^8$ photons /s /m$^2$ in Mg-K$\alpha$ and Al-K$\alpha$ at the entrance of the vacuum chamber, we will see below that for larger baselines, as the coherent part of the light is smeared out over a larger area, the build-up of an interference pattern may take several hours. For 7 keV the source strength is taken 10 times larger, to reflect the higher fluorescence yield for Fe-K$\alpha$ or Co-K$\alpha$ lines compared to Mg or Al. The spot size $S$ of these sources is presently 300 μm in diameter. Placing a slit with width $d \ll S$ in front of the source will reduce the photon flux at the entrance of the vacuum tank by a factor $E \cong Sd /(¼\pi S^2)$.

The fraction of photons that falls in the coherent part of the Fraunhofer diffraction pattern is equal to:

$$\frac{\int_{-\frac{\pi}{2}}^{+\frac{\pi}{2}} sinc^2(x)dx}{\int_{-\infty}^{+\infty} sinc^2(x)dx} \cong \frac{2.43}{\pi} \cong 0.773$$

Of these photons, a fraction $A$ of approximately 0.5 x 2 x $w$ / $D$ enters the entrance ports of the interferometer. Here the factor 0.5 accounts for the fact that the Fraunhofer pattern relative intensity drops from 1 in the center to 0.4 at the edge of the coherent pattern where the entrance ports are, hence the intensity there is roughly half the average intensity in this central part. The width $w$ of the entrance ports is of the order of 100 µm, the height is taken to be 12 mm (compatible with the size of the XCC [9]). After the entrance port are two mirror reflections with a finite transmission (cf. Table 1). The bottom row of Table 4 shows the resulting counts per second on the detector for various scaled versions of the testbed.

In order to be as insensitive as possible to all kinds of mechanical and acoustical disturbances, it seems prudent to build up the interference pattern on the sensor array as quickly as possible. As a minimum, we take ~$10^3$ photons per interference pattern (resulting in a quality similar to the one shown in [4]), but to establish 90% fringe visibilities to a sufficient degree of confidence, at least an order of magnitude more photons are needed.

The intensity required from the source scales as $D^2$: a larger baseline requires a narrower slit (limiting the emission fraction $E \propto d \propto D^{-1}$ of photons emitted by the source that pass the slit), and the photons that emerge from the slit are spread out over a larger area of which the entrance ports make up a smaller fraction (limiting the admission fraction $A \propto D^{-1}$).

Table 4. Summary of main testbed design parameters, stability requirements and source parameters for a sequence of scaled models, using the size of the vacuum tank and current Mg and Al Kα sources at Panter as boundary conditions, and the pnSensor XCC detector. Count rates are computed for a single pair of beams on a single slat on M2

| parameter | unit | scale = 0.01 | | scale = 0.1 | | scale = 1 | | | |
|---|---|---|---|---|---|---|---|---|---|
| photon energy | | 1 keV | 7 keV | 1 keV | 7 keV | 1 keV | 7 keV | 1 keV | 7 keV |
| baseline $D$ | m | 0.01 | 0.0025 | 0.1 | 0.025 | 1.0 | 0.25 | 1.0 | 0.25 |
| equiv. ang. res. | mas | 12.8 | 7.3 | 1.28 | 0.73 | 0.128 | 0.073 | 0.128 | 0.073 |
| photon spatial res. $\Delta p$ | µm | 12 | 12 | 12 | 12 | 12 | 12 | 10* | 10* |
| # pixels per fringe | | 4 | 2 | 4 | 2 | 4 | 2 | 4 | 2 |
| detector size ⊥ fringe | mm | 12 | 12 | 12 | 12 | 12 | 12 | 12 | 12 |
| grazing angle $\theta_g$ | mrad | 34.9 | 8.7 | 34.9 | 8.7 | 34.9 | 8.7 | 34.9 | 8.7 |
| min. fringe spacing $\Delta f$ | µm | 48 | 24 | 48 | 24 | 48 | 24 | 40 | 20 |
| beam angle $\theta_b$ | µrad | 25.8 | 7.4 | 25.8 | 7.4 | 25.8 | 7.4 | 31 | 8.9 |
| beam width $w$ | µm | 295 | 84 | 278 | 80 | 105 | 30 | 127 | 36 |
| # visible fringes $N_f$ | | 6.1 | 3.5 | 5.8 | 3.3 | 2.3 | 1.3 | 3.3 | 1.9 |
| overlap length $L$ | m | 11.43 | 11.43 | 10.76 | 10.76 | 4.08 | 4.09 | 4.25 | 4.24 |
| proj. arm length $B$ | m | 0.072 | 0.072 | 0.72 | 0.72 | 7.17 | 7.16 | 7.15 | 7.16 |
| total length $X$ | m | 11.5 | 11.5 | 11.5 | 11.5 | 11.5 | 11.5 | 11.5 | 11.5 |
| focal length $F$ | km | 0.387 | 0.338 | 3.87 | 3.39 | 38.7 | 33.9 | 32.2 | 28.2 |
| min. mirror size $s_m$ | mm | 8.4 | 9.7 | 7.9 | 9.0 | 3.1 | 3.6 | 3.8 | 4.3 |
| max. mirror offset $\delta x$ | nm | 36.4 | 83.1 | 36.4 | 83.1 | 36.4 | 83.1 | 36.4 | 83.1 |
| max. mirror offset $\delta y$ | nm | 1.27 | 0.73 | 1.27 | 0.73 | 1.27 | 0.73 | 1.27 | 0.73 |
| max. rms roughn. $\sigma_m$ | nm | 0.75 | 0.43 | 0.75 | 0.43 | 0.75 | 0.43 | 0.75 | 0.43 |
| max. rotation $\delta\theta$ | nrad | 298 | 149 | 298 | 149 | 298 | 149 | 248 | 124 |
| max. rel. offset $\delta m$ | nm | 2.50 | 1.43 | 2.36 | 1.35 | 0.94 | 0.53 | 0.94 | 0.53 |
| flux from source** | cts/s/m² | $10^8$ | $10^9$ | $10^8$ | $10^9$ | $10^8$ | $10^9$ | $10^8$ | $10^9$ |
| source slit width $d$ | nm | 8060 | 4606 | 806 | 461 | 81 | 46 | 81 | 46 |
| emission fraction $E$ | | 0.026 | 0.015 | $2.6\ 10^{-3}$ | $1.5\ 10^{-3}$ | $2.6\ 10^{-4}$ | $1.5\ 10^{-4}$ | $2.6\ 10^{-4}$ | $1.5\ 10^{-4}$ |
| admission fraction $A$ | | $3.5\ 10^{-4}$ | $4.0\ 10^{-4}$ | $3.3\ 10^{-5}$ | $3.8\ 10^{-5}$ | $1.3\ 10^{-6}$ | $1.5\ 10^{-6}$ | $1.5\ 10^{-6}$ | $1.8\ 10^{-6}$ |
| double reflectivity $R^2$ | | 0.42 | 0.62 | 0.42 | 0.62 | 0.42 | 0.62 | 0.42 | 0.62 |
| photon detection rate | cts / s | 385 | $3.7\ 10^3$ | 3.6 | 35.2 | 0.014 | 0.14 | 0.017 | 0.17 |

*) Extrapolated from current performance
**) Unobstructed uniform photon flux, at the entrance of the vacuum chamber

## 5.2 The testbed

Table 4 shows one possible realization of a sequence of testbeds, where the boundary conditions are taken to conform to the Panter facility: in particular the distance $r$ between the slit and the entrance ports, and the total length $X$ of the testbeds. The assumptions about the X-ray detector in this Table are based on the pnSensor XCC. The sequence shows a scaling of the baseline in steps of 10, from 1 cm for 1 keV and 0.25 cm for 7 keV up to 1 m and 0.25 m respectively. This is certainly not the only aspect that will change in the testbeds, as they also function as technology demonstrators:

- The first testbed will most likely be build on a regular stainless steel optical bench. With the small baseline, the countrate is high enough to build up a fringe pattern in mere seconds, and the optical path is fully dominated by the beam overlap region, which means that thermal instability is probably not a limiting factor, compared f.i. to mirror quality. Also, the source exit slit is of the order of several micron, which can still be realized with a knife edge, rather than a lithographically fabricated slit.
- As was noticed earlier: the stability requirements do not depend on baseline, and are therefore the same for all testbeds. What does change is the distance over which they need to be maintained, as the baseline grows and the beam overlap length $L$ shrinks, which make them harder to maintain, both in $x$ and $y$.
- For the next models the count rates are already a factor 100 lower, so here we may have to consider either additional thermal stabilization, possibly aided by metrology to facilitate localization of the fringes, and / or the use of SiC as bench material, as the thermal stability is likely to come into play.
- An increase of the source strength is another avenue to pursue: a stronger source would offer an opportunity to decouple the effect of thermal stabilization form other aspects that might impact on the fringe visibility. The source spot of the current sources is 300 μm in diameter, which means that a large fraction of the generated photons is thrown away by placing a narrow slit in front of the source. Reducing the source spot without reducing the flux, e.g. by using small focusing optics, would clearly help. The realization of one of the more advanced models at a synchrotron facility might be an alternative option.
- As the required slit width falls below 1 micron, slits need to be produced lithographically. Techniques for producing high-aspect ratio slits need to be developed in parallel to the testbeds.
- Another aspect that needs further development is the accuracy with which the photon absorption site can be determined, or the sub-pixel resolution, $\Delta p$. This is in particular relevant for the 7 keV interferometry. As Table 4 shows, even with a quarter of the full baseline, and two pixels per fringe the number of fringes per beam width is not even two for the largest scale testbed. There are at least two possible avenues for improvement that can be explored (other than resorting to smaller baselines for 7 keV photons)
  - Tilting the camera out of plane under an angle $\alpha$, so that the interferogram is spread over an area that is a factor $\cos^{-1}\alpha$ larger. In order to increase the number of fringes by a factor of 2 or more, a tilt angle $\alpha > 60°$ is required.
  - Changing the shape of pixels, or the electrodes that define the pixel shape, into a rectangular shape: high position definition is only needed in the direction of the fringes. The challenge is to maintain a sufficiently good energy resolution to not reduce the fringe visibility.

  As can be seen in Table 4 (shaded cells), a small improvement in $\Delta p$ already makes a more than linear proportional difference at 7 keV.
- Table 4 lists count rates for a single pair of interferometric beams. But an essential part of the design shown in Figure 1 is the slatted mirror, M2, which acts as a multiplexer, in the sense that it allows multiple parallel beams to interfere. This is also a route to improve the count rate by a factor of 10 or more and hence the build-up time for the interferogram in the follow-up models. The viability of the slatted mirror concept has already been demonstrated for optical wavelengths [7]. The demonstration of multiplexed interferometry, and the associated alignment is an important part of the technology demonstration, and will have to be introduced early on.

## 6. SUMMARY AND CONCLUSIONS

We have identified a single spacecraft X-ray interferometer concept capable of 30 - 60 µas angular resolutions in the astrophysically relevant energy range 1 – 7 keV. In a basic stability requirements analysis we found no show-stoppers: on the contrary, the currently flying ESA mission GAIA achieves optical stabilities two orders of magnitude more stringent than appear to be required for an XRI. Even if we err on the optimistic side in this study there appears to be quite some margin before we hit the limits of technology.

We have outlined a series of testbed experiments to raise the technological maturity of X-ray interferometry from laboratory demonstrations with baselines at or below a mm to a representative bench capable of interferometry in the energy range 1 – 7 keV with baselines of the order of 1 meter. In particular, the first model appears to be feasible with existing technology, while with improvements on this technology the succeeding models can be realized.

## 7. ACKNOWLEDGMENTS

We acknowledge informative conversations with Max Colon from cosine BV, Warmond, the Netherlands, Lothar Strüder of PNSensor GmbH, München, Germany, and Vadim Burwitz, of the PANTER X-ray Test facility, Max-Planck-Institut für Extraterrestrische Physik, Neuried, Germany.